\documentstyle[amssymb,aps,preprint]{revtex}

\begin{document}

\begin{center}
{\LARGE Growth kinetic study of Sputter Deposition: }

{\LARGE Ag on Si/SiO}$_{2}${\LARGE .}
\end{center}

\bigskip

\begin{center}
S. Banerjee and S. Kundu.

Surface Physics Division, Saha Institute of Nuclear Physics,

1/AF Bidhannagar, Calcutta 700064, India.
\end{center}

We have presented the kinetic study of the very initial growth stages of an
ultra thin film (40\AA\ - 150\AA ) of Ag sputter-deposited on Si(001)
substrate containing native oxide using grazing incidence x-ray reflectivity
(GIXR) technique and Atomic Force Microscopy (AFM). We observe that the film
consists of mounds with the presence of voids. The thickness `$d_{xray}$'
and the packing fraction `$\eta $' of the film as a function of time `$t$'
follow a simple power law, $d_{xray}$ $\thicksim $ $t^{m}$ and $\eta $ $%
\thicksim $ $t^{n}$ with the exponent $m$ = 0.58 and $n$ = 0.37
respectively. We have quantitatively determined that the voids between the
mounds decrease at the initial growth stages with the increase in mound size
using GIXR measurement. The mound size increases mainly by the coalescence
process on the substrate. We have observed that as a function of time the
mound size $R(t)$ increases radially as $\thicksim $ $t^{z}$. The radial
growth exponent $z$ crosses over from $z$ $>$ 0.5 to $z$ $\thicksim $ 0.25
indicating two growth regimes. The GIXR measurement reveals sublinear
dependence of $\eta $ on $d$ and the AFM measurement shows a cross over of
the radial growth exponent, both these indicates that the lateral growth of
the mound is enhanced initially reducing the voids.

\bigskip


e-mail: sangam@cmp.saha.ernet.in

\newpage

The study on the growth of ultra thin metallic films are of great importance
not only for understanding the growth kinetics at its fundamental level but
also for its technological application in the fabrication of nano-devices.
In the semiconductor industries, sputtering is commonly used for depositing
ultra thin films for metallization. Here we report a study on the growth
kinetics of the initial growth stages of Ag films using grazing incidence
x-ray reflectivity (GIXR) technique \ and Atomic Force Microscope (AFM).
GIXR technique has been successfully used in a non-destructive manner to
obtain the structure and chemical profile of the film as a function of its
depth \cite{PRB1,JAP,PRB2}. Atomic force microscope (AFM) is used for
obtaining topographic images of hard and soft surfaces with Angstrom
resolution. Recently we have reported that when Ag is deposited using dc
magnetron sputtering, the initial growth of the film is carpet-like and
above a critical thickness $\thicksim $ 40 \AA\ the film structure changes
to form mounds \cite{jpd2}. Below the critical thickness ( $\thicksim $ 40 
\AA ) one can easily modify the ultra thin film of Ag intentionally by the
tip of AFM to fabricate periodic pattern of nanoscale order \cite{jpd2,epjap}
and above the critical thickness of the film one observes the onset of mound
formation and these mounds are significantly stable and cannot be modified
by the AFM tip. The morphological shape of the mound could be determined by
combining AFM and GIXR measurement \cite{jpd1}.\ In the present
investigation we have studied the growth kinetics of ultra thin Ag film and
the coarsening process of the mounds by following its evolution with respect
to the deposition time using ex-situ AFM and GIXR measurement.

Formation of mounds in the initial growth stages of thin films of various
materials have been observed earlier \cite
{Mark,Joseph,Lengel,Georgios,Je,Jeffries,Karr,You} and extensive theoretical
work \cite{Martin123,Rost,Amar} have been carried out in particular for
understanding the radial coarsening law $R(t)$ $\thicksim $ $t^{z}$ for the
increase of the typical mound size $R$ with respect to time `$t$' which are
observed in experiments \cite{Mark,Joseph,Lengel,Georgios,Karr} and
simulations \cite{Smilauer,Johnson}. Most of the experimental work reported
to our knowledge is based on film having thickness greater than the present
study \cite{Mark,Joseph,Lengel,Georgios,Je,Jeffries,Karr,You} and also grown
using molecular beam epitaxy (MBE) \cite{Mark,Joseph,Lengel,Georgios}. Only
a few studies on the kinetics of mound growth of sputter deposited Pt \cite
{Jeffries}, Au \cite{You}, Ag \cite{Je} and TiN \cite{Karr} films have been
reported earlier. The kinetics of the growth of the mounds can be different
in the case of films deposited using magnetron sputtering than that grown
using MBE. The present investigation will be of interest for comparing the
growth of the sputter deposited film with that grown using MBE. In this
letter we would like to address mainly the issue of time dependent
coarsening of the mound using GIXR and AFM\ measurement. We have observed
that mound size increases with time mainly by coalescence process and tried
to look into the kinetic of its growth. We would also like to emphasise here
that the film grown by sputtering techniques exhibits similar exponents only
after certain initial stage of the film growth\ as predicted by certain
theories \cite{Martin123,Rost,Amar} and by the films grown by MBE \cite
{Mark,Joseph,Lengel,Georgios}. At the very initial stages of the film growth
by sputtering technique we observe a higher value of growth exponent.

For the present work we have deposited thin films of silver (99.99 $\%$,
Target Material Inc.) on Si(001) substrate using a dc magnetron sputtering
unit (Pfeiffer, PLS500). The sample preperations and deposition parameters
are given in our earlier work, see reference \cite{jpd2,epjap}.\ Six
different films were\ deposited as a function of time (t = 20, 30, 40, 60,
120, 180 secs.) for obtaining various thicknesses labeled as sample (a) to
sample (f). The film thicknesses were measured using x-ray reflectivity
technique \cite{Born}. For x-ray reflectivity measurement we have used $%
\theta -2\theta $ diffractometer (Microcontrol Inc.). The Cu-K$_{\alpha 1}$
x-ray was obtained from 18KW rotating anode x-ray generator (Enraf Nonius
Inc.). For obtaining the surface morphology of the film we have used atomic
force microscope (Park Scientific Inc.). The scan was carried out using the
AFM in contact mode and the z-calibration of the piezo was done using x-ray
reflectivity measurement \cite{rsi}. The measurements were carried out in
air at room temperature.

We have used x-ray reflectivity technique to characterise the thickness and
average electron density (AED) of the films. The AED of the film is related
to mass and volume density of the film \cite{PRB1}. The x-ray reflectivity
data has been presented by the authors recently \cite{jpd2,epjap}, but the
following analysis has not been presented earlier. The thickness 'd$_{xray}$%
' and the packing fraction '$\eta $' obtained from the x-ray reflectivity
data analysis are plotted as a function of deposition time '$t$' and are
shown in fig.1(a) and 1(b) respectively. Both follow a simple power law
relation $d_{xray}$ $\thicksim $ $t^{m}$ and $\eta $ $\thicksim $ $t^{n}$
with the exponent $m$ = 0.58 and $n$ = 0.37 respectively, obtained from the
non-linear least square fitting (shown as solid line in fig. 1). The lower
value of the exponent for the case of $\eta $\ implies that at the very
initial growth stages the increase of packing fraction is more dominant than
the increase of thickness as a function of time indicating that the
deposited particle tries to spread laterally reducing the {\it voids} much
faster than growing vertically for increasing the thickness of the film. A
linear fit to the data plotted with its logrithmic values are also shown in
the insets for better clarity. From the previous two power laws one can
write a relation between $\eta $ and $d$ as $\eta $ $\thicksim $ $d^{p}$
where $p=n/m$. Using the values of $m$ and $n$ obtained from the fits, $p$ =
0.64 and fig 1(c) shows this plot. The sublinear dependence of $\eta $ vs $d$
also indicates that the packing fraction grows faster initially than the
thickness. Thus in the sputter deposition process of Ag on Si/SiO$_{2}$ the
mound grows more rapidly in the lateral direction initially than
vertically.\ The above growth mechanism indicates that the growth of the
film by sputter deposition at its initial growth stages is very similar to
Stranski-Krastanov type of growth \cite{Schmidt}.

AFM images for the six different samples and the typical line profiles for
each of the images are shown in fig. 2 and fig. 3 respectively. The grain
''mound'' size can be estimated from the line profiles. We observe from the
AFM image (fig.2a) and from the line profile (fig. 3a) that for sample (a)
the mounds are monodispersed. As the time of deposition increases the mound
size increases as marked by arrows in the line profile and labeled as ''A''
in the AFM images and in the line profiles. The number density of the mound
decreases as the growth of the film continues, this indicates that the mound
grows by coalescence process. Some ``A'' types mounds coalesce to form
bigger mounds and are labeled as ''B''. As the deposition progresses further
we observe in the case of sample (f) that there are now mainly three types
of mound sizes and are marked as A, A1 and B in the AFM image. A1 is an
intermediate-size mound, formed by the coalescence of mounds with fewer
nearest neighbours. Many such intermediate size of mounds will develop from
this instant and this makes us difficult to keep the track of the mounds
size as a function of time and thus we have stopped the growth beyond this
time. This indicates, that as the time progresses the size distribution
changes from mono-dispersed to polydispersed. We are concentrating in the
present work only on the initial growth stages and hence we have\ selected
mainly the two types of mounds A and B. We have plotted the mound size of
type A and type B as a function of time in fig. 4. We observe that as a
function of time the mound size $R(t)$ increases as $\thicksim $ $t^{z}$,
where the radial growth exponent $z$ crosses over from a linear regime $z$ $%
\thicksim $ 1 to $z$ $\thicksim $ 1/4 for the mound type A and for the mound
type B the exponent $z$ crosses over from $z$ $\thicksim $ 0.7 to $z$ $%
\thicksim $ 1/4 at the same instant of time, indicating that there are two
growth regimes during the coarsening process for the film grown using
sputtering technique. The cross over of the exponent for both the cases
occurs at the same time i.e., at time t = 60 secs. (This time depends on the
growth conditions). The higher value of the exponent at small time 't'
indicates that the mound growth in the radial direction is enhanced and
beyond time t = 60 secs the radial growth is suppressed. The enhanced radial
growth region and the suppressed radial growth region are shown in fig. 4.
Thus, the analysis of the AFM results give similar interpretation as that
obtained from the analysis of the x-ray reflectivity data. We can also infer
from the above analysis that the growth of both the type of mounds (type A
and type B) follows more or less a similar growth exponent (i.e., for t 
\mbox{$<$}%
60s the growth exponent z 
\mbox{$>$}%
0.5 and for t 
\mbox{$>$}%
60s the value z 
\mbox{$<$}%
0.5 for both the type of mounds).

We now look into the scaling aspects during the dynamic growth process. We
have determined the rms roughness $w(t)=\langle \lbrack h(r,t)-\langle
h\rangle ]^{2}\rangle ^{1/2}\varpropto t^{\beta }$ for $r>>\xi (t)$. The
average $\langle ...\rangle $ is the spatial average over the sample
surface, $h$ is the height on the surface seperated by a lateral distance $r$
and $\xi (t)$ is the lateral correlation length which is a measure of the
average mound size. The $\xi (t)$ $\thicksim $ $t^{z}$, where $z=\beta
/\alpha $, $\beta $ being the growth exponent and $\alpha $ is the roughness
exponent \cite{Barabasi}. In the inset of fig. 4 we have plotted $w(t)$ vs $%
t $ in a log-log scale and we again observe the power law dependence as $%
w(t) $ $\thicksim $ $t^{\beta }$ where $\beta $ $\thicksim $ 0.75 for $t<60s$
and $\beta $ $\thicksim $ 0.15 for $t>60s$. This also clearly indicates that
there are two different regimes of mound growth (Beyond t = 60s we observe
reduction of rms roughness). Recent theoretical work predicts that the
characteristic length $\xi (t)$ scales as $\xi (t)$ $\thicksim $ $R(t)$ $%
\thicksim $ $t^{z}$ where $z$ $\thicksim $ $1/4$ \cite{Martin123,Rost,Amar}.
We see from fig. 4 that $R(t)$ follows the theoretical prediction for $t>60s$
only. For the very initial growth stages which has not been studied
carefully both experimentally and theoretically, we observe that the radial
(mound) growth exponent exhibits a higher power law. Thus, at the initial
growth stages we observe a cross over of the radial growth exponent
indicating that when the substrate surface is fully covered up by the mounds
supressing the substrate effects the lateral growth of the mound is also
supressed. We would like to make a remark here that we have not used the
roughnesses obtained from the specular x-ray reflectivity for the above plot
of $w(t)$ vs $t$ (inset of fig. 4). Recently one of the author has pointed
out that the parameter $\sigma $ \cite{Nevot} which is the FWHM of of the
gaussian distribution of the derivative of the electron density `$d\rho
(z)/du$' where $\rho (u)$ is the electron density as a function of depth $u$
of the film which most often is termed as roughness may not be the true
roughness $w$ as defined above \cite{JAP2}.

In conclusion, we have presented a comprehensive picture of film growth by
sputtering deposition. We have followed the growth kinetics of the initial
growth stages of Ag film grown by sputter deposition using GIXR and AFM
measurements. We have proposed coalescence process as the growth mechanism
for the increase of mound size as a function of time during the growth of
the film. We have observed cross over of the radial growth exponent of the
mounds indicating that there exist two distinct growth regimes. As a
function of time during the growth of the film, initially the mound in the
film exhibits enhanced lateral growth showing higher value of radial growth
exponent and as the time progresses the lateral growth is suppressed. The
two regimes have different kinetics of growth. The mound growth exponent $z$ 
$\thicksim $ 1/4 agrees with the theoretical and the simulation prediction
for the later stages of growth when the substrate effect is supressed.

\bigskip \newpage

\begin{center}
{\Large Figure Captions:}
\end{center}

\bigskip

FIG. 1. Time evolution of (a) thickness `$d$', and (b) packing fraction `$%
\eta $'\ for the sputter deposited Ag films. The solid lines shows that both
follows a simple power law, $d_{xray}$ $\thicksim $ $t^{m}$ and $\eta $ $%
\thicksim $ $t^{n}$ with the exponent $m$ = 0.58 and $n$ = 0.37
respectively. The packing fraction $\eta $ is related to $d$ as $\eta $ $%
\thicksim $ $d^{p}$ where $p=n/m=$ 0.64, the solid line shows the plot. In
insets we have plotted the logrithmic values and we show the linear fit for
better clarity.

\bigskip

FIG. 2. AFM images for sample (a) to sample (f). As the time of deposition
increases the mound size increases by coalescence process labeled as ``A''.
Some A type mounds coalesces to form mounds of type B and A1 are the mound
at later stages. The solid lines in the images are the marker for the\ line
profiles and the respective arrows are shown in FIG. 3.

\bigskip

FIG. 3. Typical line profiles for sample (a) to sample (f) marked as solid
lines in the AFM images of FIG. 2. The arrows indicated by label ``A'' are
for mounds of type A and label ``B'' are for mounds of type\ B.

\bigskip

FIG. 4. The time evolution the radial mound size $R(t)$ increases radially
as $\thicksim $ $t^{z}$. The radial growth exponent $z$ crosses over from a
linear regime $z$ $\thicksim $ 1 to $z$ $\thicksim $ 1/4 for mound A and
from $z$ $\thicksim $ 0.7 to $z$ $\thicksim $ 1/4 \ for mound B at the same
instant of time $t=60s$ indicating two growth regimes. The two growth
regimes are marked by a vertical dash line seperating the enhanced lateral
growth regime from the suppressed lateral growth regime. The inset shows the
rms roughness $w(t)$ $\thicksim $ $t^{\beta }$ where the growth exponent $%
\beta $ $\thicksim $ 0.75 for $t<60s$ and $\beta $ $\thicksim $ 0.15 for $%
t>60s$ showing different values in both the regimes.

\end{document}